\documentclass{mem}
\usepackage{natbib}
\usepackage{txfonts}
\usepackage{balance}
\usepackage{flushend}
\usepackage{graphicx}
\usepackage[breaklinks,pdftex]{hyperref}
\usepackage{aas_macros}
\usepackage{xcolor}
\idline{75}{282}
\begin{document}
\def\teff{$T\rm_{eff }$}
\def\kms{$\mathrm {km s}^{-1}$}

\title{
Cosmic-ray propagation features in gamma-ray measurements
}

   \subtitle{}

\author{
J.\ Becker Tjus\inst{1,2,3} 
\and J.\ D\"orner\inst{1,2} 
\and L.\ Schlegel\inst{1,2}
\and W.\ Rhode\inst{1,2,4}
          }

\institute{
Ruhr University Bochum, Faculty for Physics and Astronomy, Institute for Theoretical Physics IV: Plasma-Astroparticle Physics, 44780 Bochum, Germany
\and
Ruhr Astroparticle and Plasma Physics Center (RAPP Center), 44780 Bochum, Germany
\and 
Department of Space, Earth and Environment, Chalmers University of Technology, 412 96 Gothenburg, Sweden
\email{julia.tjus@rub.de}
\and
TU Dortmund University, Faculty for Physics, 44221 Dortmund, Germany\\
}

\authorrunning{Becker Tjus \textit{et al.}}

\titlerunning{Constraining cosmic-ray propagation with gamma rays}

\date{Received: XX-XX-XXXX; Accepted: XX-XX-XXXX}

\abstract{
$\gamma$-ray measurements from GeV to PeV energies have provided us with a wealth of information on diffuse emission and sources in the Universe lately. With improved spatial and temporal resolutions together with real-time multimessenger astronomy, the modeling of 3D cosmic-ray transport becomes more and more important to explain the data. Here, we will give a compact summary of how cosmic-ray propagation in very different astrophysical environments like the Sun, Milky Way, and active galaxies can be constrained by combining 3D modeling with the propagation software CRPropa with $\gamma$-ray measurements.
}
\maketitle{}
\parindent=0cm
\section{Introduction}
New instruments in $\gamma$-ray astronomy have boosted our understanding of the high-energy Universe. The Fermi satellite is cataloguing the Galactic and extragalactic skies at GeV energies with previously unknown sensitivity. The second generation IACTs (H.E.S.S., MAGIC, VERITAS), together with the Water Cherenkov Arrays (MILAGRO, HAWC, LHASSO) have boosted our understanding of the non-thermal processes from TeV to PeV energies. Due to new instruments and techniques, sources and source regions can now be resolved in space and time with high precision. This wealth of data calls for quantitative models that put the $\gamma$-ray measurements in the multimessenger context. 
The requirements for the mulimessenger codes are very different
whether the description is to be done for example for the full Milky Way \citep{Galprop_Web_Standford,dragon2,picard2014},
or highly variable sources like blazars \citep[e.g.][]{a7:dimitrakoudis2012,boettcher2013,a7:gao2017,reimer2023}.  Here, we will present the modular propagation tool CRPropa, which is adjustable to a larger set of source environments \citep{CRPropa3.1,CRPropa3.2} and discuss a selection of sources - the Sun, the Milky Way, and active galaxies - for which CRPropa (or a modification of the original code) can be used to help understand where the different emissions come from and what we can learn about cosmic-ray (CR) propagation from these results.




\section{Cosmic-ray propagation and interactions with CRPropa 3.2}
%
CR propagation and interaction in galactic and extragalactic environments can be modeled within the CRPropa framework \citep{CRPropa3.2}. Its modular structure allows easy customization for different astrophysical applications. Every physical process is implemented as an independent module, and the user can decide which modules should be included in the simulation. 

The latest public version 3.2 offers several relevant interaction channels for CR nuclei, CR electrons, and gamma rays, where different photon-field models for the extra-galactic background light (EBL) as well as the CMB \cite{CRPropa3.2}. For other applications like in-source propagation or Galactic CRs, the interaction rates for a user-defined photon can be pre-computed. Several plugins to extend the implemented physical processes are available beyond the official version.  \citet{Doerner25} implements the proton-proton interactions leading to the production of secondary $\gamma$-rays and neutrinos. An overview of available plug-ins is collected on the CRPropa documentation\footnote{\url{https://crpropa.github.io/CRPropa3}}.

The first available option in CRPropa for CR propagation is the solution of the equation of motion, which is possible for arbitrary background magnetic fields. 
 CRPropa offers different coherent magnetic field models like the Galactic Magnetic Field (GMF) as well as different descriptions of synthetic turbulence either on a grid or using a plane-wave method. The second option is the solution of the diffusion-advection equation 
\begin{equation}
    \partial_t n = \nabla(\hat{\kappa}\nabla n) - \vec{u} \nabla n \,,
\end{equation}
using the method of stochastic differential equations \citep[SDEs; ][]{CRPropa3.1}. Here, $\vec{u}$ describes the bulk motion of the background plasma. The spatial diffusion can be anisotropic with respect to the local background magnetic field. The diffusion tensor is $\hat{\kappa} = \mathrm{diag}(\kappa_\perp, \kappa_\perp, \kappa_\parallel)$ assuming $\vec{B} = B_0 \vec{e}_z$. 

With these two different propagation mechanisms, CRPropa offers a unique way to test the transition between the diffusive regime at lower energies and the ballistic behavior at higher energies within a consistent framework. In the following, we will discuss different applications of the CRPropa framework.

\section{Gamma-rays from the Sun: hadronic emission?}


The Sun is the closest source of very high-energy gamma rays and is detected in the GeV range by FermiLAT \citep{Fermi_Sun_11, Tang18, Linden18, Linden22} and in the TeV range by HAWC \citep{HAWC_Sun}. Those gamma rays are believed to originate from hadronic interactions of Galactic cosmic rays (GCRs) within the solar atmosphere \citep{Seckel-etal-1991}. 
Due to the high Lorentz boost at the production of the secondary gamma rays, their production is mainly in the direction of the primary proton. As most protons are directed toward the solar surface, even most of the produced gamma rays are directed that way and will therefore be absorbed. To observe gamma rays from the Sun, the primary CRs must be reflected in the solar magnetic field before the interaction happens, or the interaction must occur far enough away from the solar surface in a way that absorption processes are negligible. 

\begin{figure}[t]
    \centering
    \includegraphics[width=\linewidth]{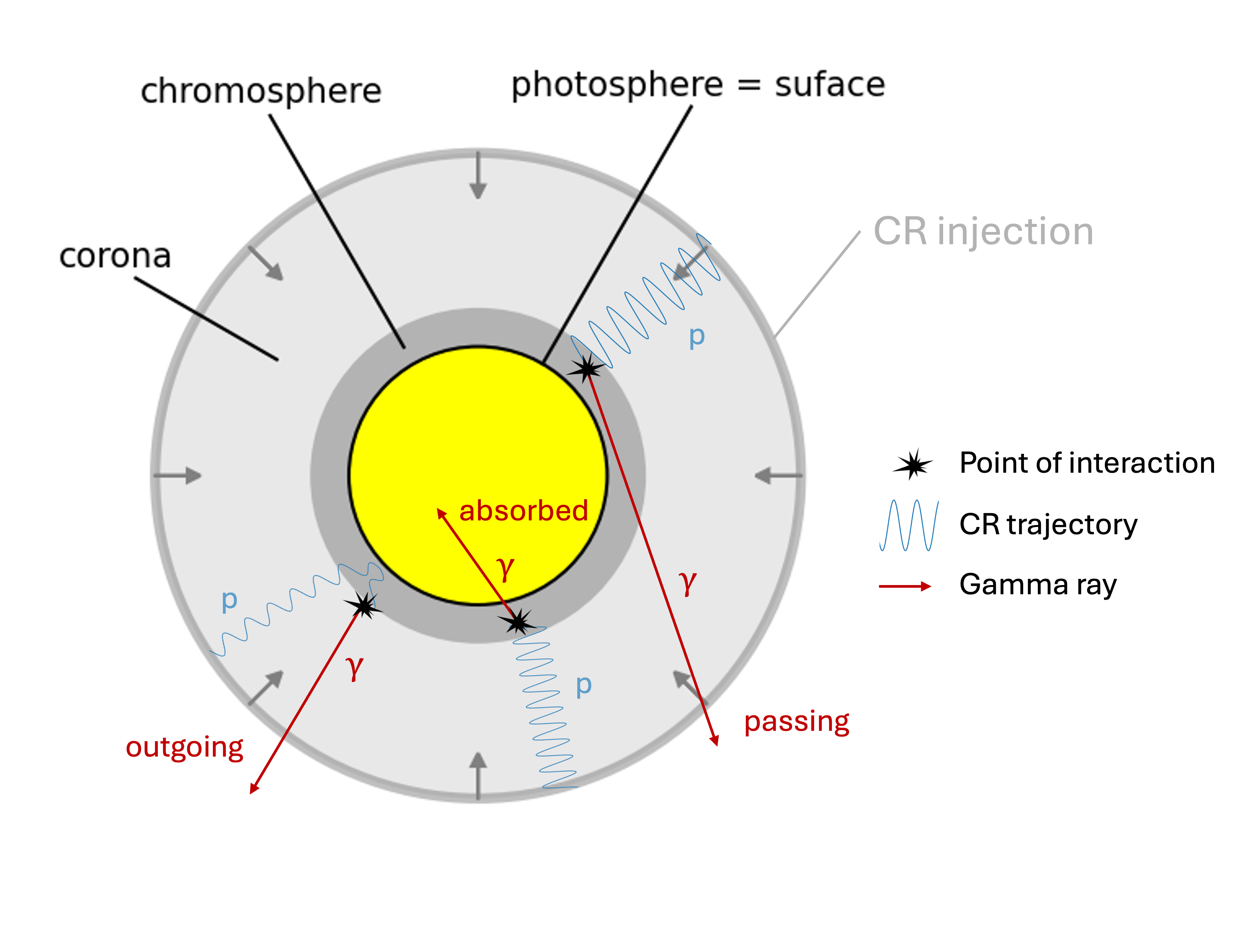}
    \caption{Schematic view of the CR trajectories in the solar atmosphere.}
    \label{fig:sol_schematic}
\end{figure}

To simulate the interaction of GCRs within the solar atmosphere we consider the injection of protons from a sphere $2.5$ solar radii away from the Sun. The GCRs are propagated ballistically through the solar magnetic field. To infer the impact of different field geometries, the potential field source surface (PFSS) model, evaluated for different Carrington Rotations, and the dipole/quadrupole and current sheet (DQCS) model introduced by \cite{Banaszkiewicz98} are used.  

A schematic overview of the different production directions and the simulation setup is given in Fig.\ \ref{fig:sol_schematic}. In the output, only those gamma rays passing the solar surface or having an outgoing direction are considered. 

\begin{figure}[t]
    \centering
    \includegraphics[width=\linewidth]{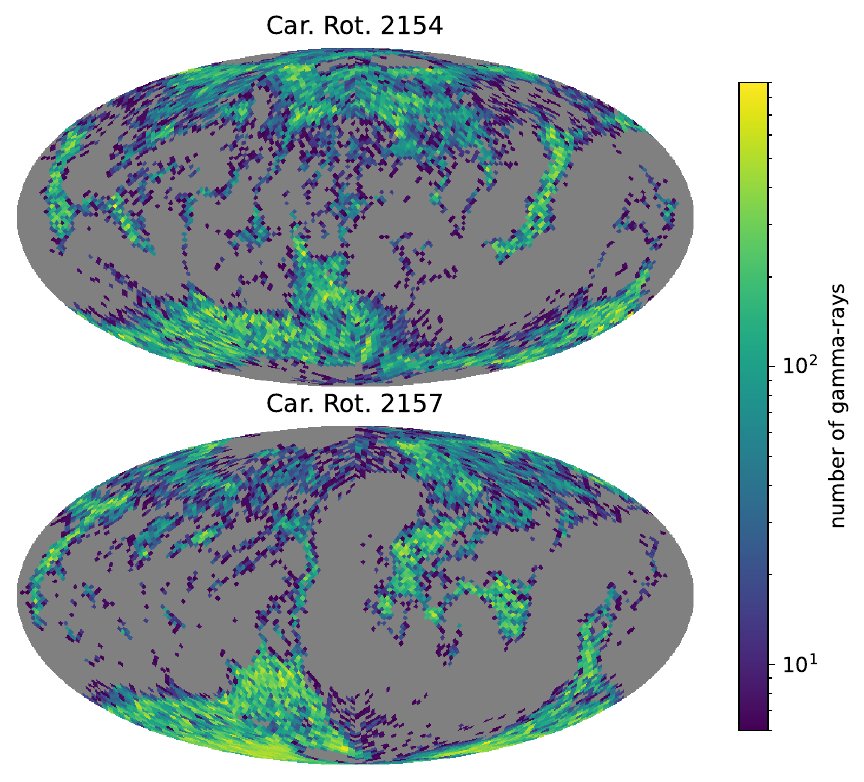}
    \caption{Spatial distribution of created $\gamma$-rays projected on the solar surface in different magnetic field geometries.}
    \label{fig:sol_maps}
\end{figure}

In the first step, the influence of the magnetic field structure on the spatial distribution of the produced gamma rays is evaluated. For this, we use a mono-energetic injection of protons with $E_p = 100$ GeV and the PFSS magnetic field for the Carrington Rotation 2154 and 2157. The projected position of the produced gamma rays on the solar surface is shown in Fig.\ \ref{fig:sol_maps}. 

In both maps, most produced gamma rays lie within the polar-cap region. This region is dominated by open field lines that connect the solar surface with the interplanetary space. The CR can follow those field lines to reach the innermost layers, where the target density is high enough to trigger interactions. The more localized structural differences between the maps are caused by solar activity and changes in the field structure. Empty regions in gamma rays correspond to closed field lines where the CR particles cannot reach the atmosphere or the sunspots, which have the highest magnetic field strength. Due to the higher strength, the mirroring point is further away from the Sun, and the particles do not traverse enough target to trigger an interaction. 

\begin{figure}[t]
    \centering
    \includegraphics[width=\linewidth]{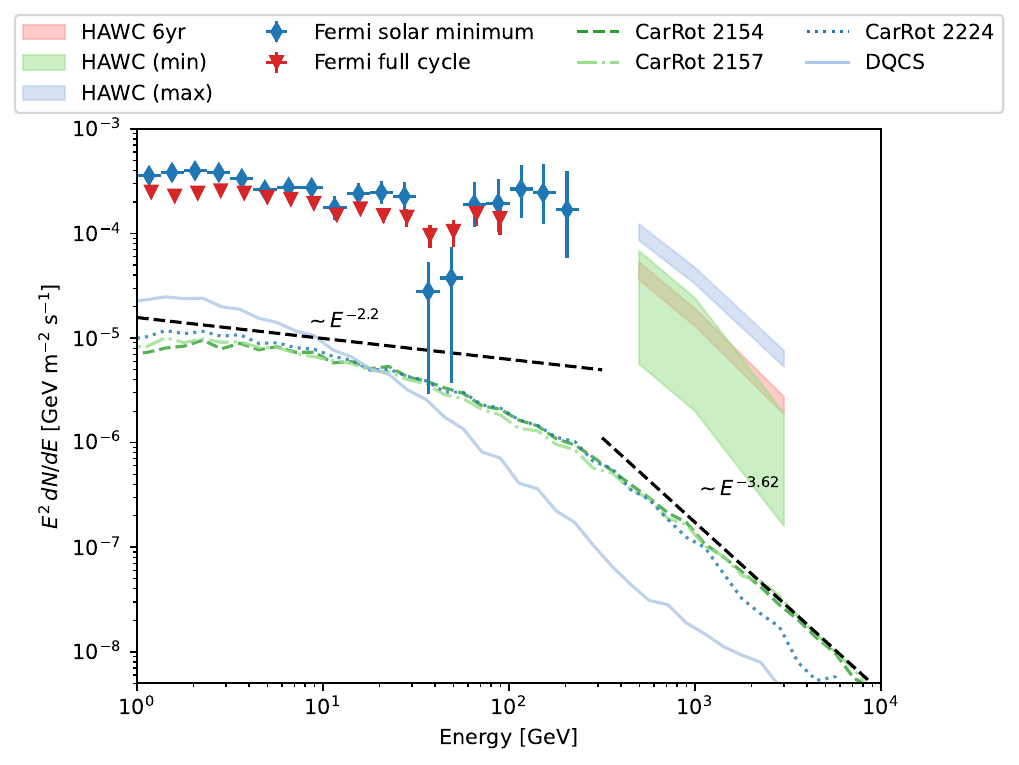}
    \caption{SED of the solar gamma rays compared with data from Fermi-LAT \citep{Linden18, Linden22} and HAWC \citep{HAWC_Sun}.}
    \label{fig:sol_spectrum}
\end{figure}

To model the full SED of the solar disk, we simulate the local interstellar spectrum as it arrives outside of the heliosphere as the starting energy distribution. This neglects the solar modulation, which is reasonable for the higher energies ($\sim$ TeV) and can be interpreted as an upper limit for the lower energies ($\sim$~GeV). The resulting flux is evaluated for three different PFSS models (Carrington Rotation 2154 and 2157 close to the solar maximum and 2224 at the solar minimum) and the DQCS model. The results are shown in Fig.\ \ref{fig:sol_spectrum} and compared to the observations by Fermi-LAT and HAWC. 

The SED shows clear differences between the DQCS and the PFSS models. The latter can reproduce the expected energy scaling in the lower ($\sim E^{-2.2}$) and higher ($\sim E^{-3.62}$) energy ranges but predicts a significantly lower $\gamma$-ray flux. The simulation does not show the seasonal variation between the solar minimum and maximum. The predicted energy scaling of the DQCS field for the highest energies matches the observation, but the total normalization is even lower than the PFSS prediction. In the lower energies, the energy scaling cannot be reproduced. This field configuration would require an additional energy-dependent transport effect. 
To explain the missing fraction of the total $\gamma$-ray flux, additional target material would be needed. \cite{Li24} propose a scenario with additional interactions in internetwork regions below the surface. In the future, it will become necessary to combine these lower layers with the large-scale structured magnetic field from the corona and to consider a more realistic target distribution in the chromosphere. 

\section{Gamma-rays from the Galactic Plane and anisotropic diffusion}
The diffusion of CRs in a magnetic field, composed of a coherent background field $\vec{B}$ and a turbulent component $\delta\vec{b}$ is expected to be anisotropic for small turbulence $\delta b^2 \ll B^2$ \citep[e.g.][]{beckertjus_merten2020}. This effect is known from analytical calculations in the quasi-linear theory and from test particle propagation in synthetic turbulence \citep{2022MNRAS.514.2658R,2022SNAS....4...15R}. 
In the modeling of Galactic CR transport, the effect of anisotropic diffusion has become of particular interest in the past years \citep{Effenberger12, Giacinti13}. 
The first step to constrain the anisotropy of the diffusion tensor with $\gamma$-ray observations has been done by \cite{Dörner24}. Here, the Central Molecular Zone (CMZ) of the Milky Way is modeled. The anisotropy of the diffusion tensor is described by the ratio between the diffusion coefficient parallel and perpendicular to the background magnetic field $\epsilon = \kappa_\perp / \kappa_\parallel$. The discrimination between the more isotropic ($\epsilon \approx 1$) and the strongly anisotropic diffusion ($\epsilon \ll 1$) is based on the spatial $\gamma$-ray distribution measured by H.E.S.S.\ \citep{Abramowski2016}. For the CMZ the best agreement to the observation is achieved using isotropic diffusion. 
In the next step, we extend the work from \cite{Dörner24} to the full Milky Way. In contrast to the work for the CMZ, a direct follow-up of the gamma rays produced in the $pp$ interactions is not feasible as it would require an intense amount of computation power. For a CR with $1$ TeV that would stay for $\tau = 1$ Myr within the Galaxy, an average target density of $n_\mathrm{H} = 0.01 \, \mathrm{cm}^{-3}$ in the galactic halo and the cross section in the order of $\sigma_{pp} \sim 10^{-26} \, \mathrm{cm}^{2}$ the expected number of interactions per primary CR can be calculated as 
\begin{equation}
    N_\mathrm{int} \approx c \tau n_\mathrm{H} \sigma_{pp} \approx 10^{-4} \, .
\end{equation}
To achieve a reasonable resolution on the sky in total $N_\mathrm{pix, sky} = 10^5$ pixel are needed. Together with an energy resolution of $N_\mathrm{pix, E} = 80$, a total amount of $N_\mathrm{sim} = 8 \cdot 10^{12}$ simulated particles would be required to result on average to a 10 \% statistical uncertainty in each bin. 

\begin{figure*}[ht]
    \centering
    \includegraphics[width=0.8\linewidth]{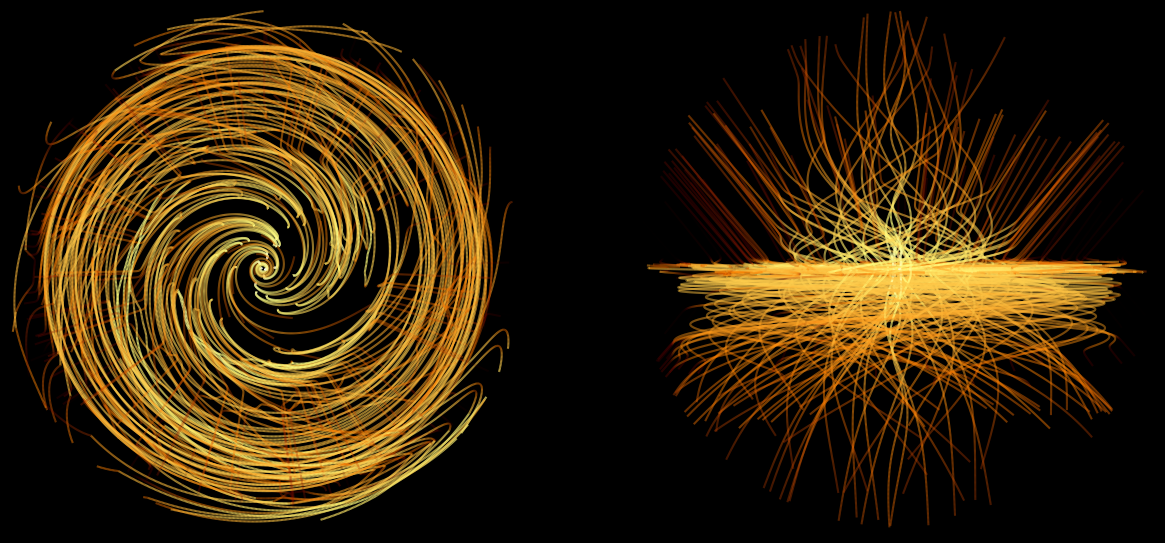}
    \caption{Magnetic field lines from the combined solenoidal JF12 model \citep{JF12_sol} and the CMZ model \citep{Guenduez2020ACenter} in the face-on (left) and edge-on (right) view. 
    }
    \label{fig:field}
\end{figure*}

\begin{figure}
    \centering
    \includegraphics[width=\linewidth]{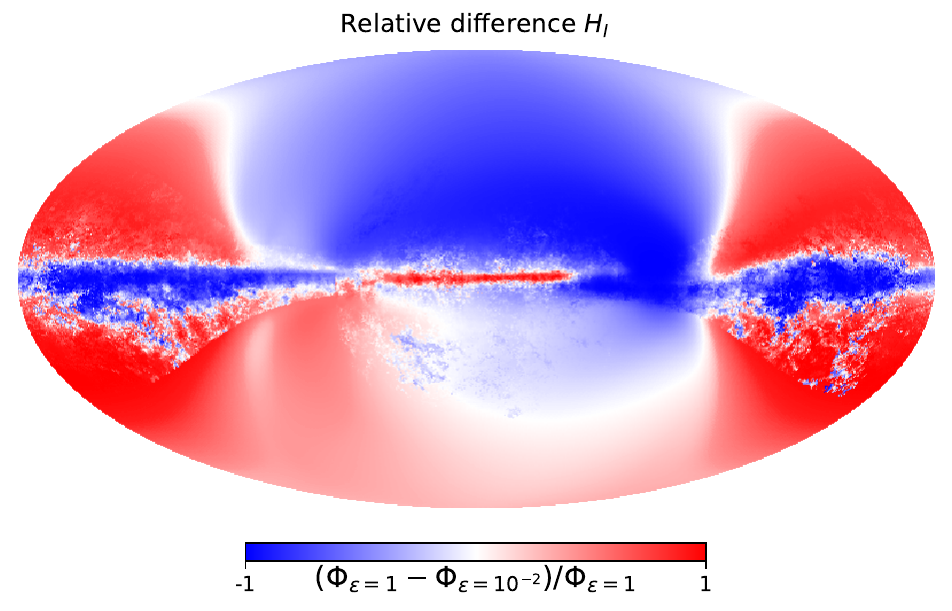}
    \caption{Relative difference between the expected $\gamma$-ray flux from $pp$ interactions on interstellar H$_I$ gas in the isotropic ($\epsilon = 1$) and anisotropic case ($\epsilon = 10^{-2}$). $E_\gamma = 100$ GeV.}
    \label{fig:skymap}
\end{figure}

For an efficient simulation, we use the CRPropa framework to calculate the 3+1 dimensional CR distribution in the Milky Way as a function of space and energy. After the CRPropa simulation, we use the HERMES framework \citep{Hermes} to calculate the total $\gamma$-ray intensity for a given sky position as the integral over the line-of-sight
\begin{eqnarray}
    I_\gamma(E) = \int\limits_{0}^{\infty} \mathrm{d}s \, n_H(\vec{r}) \int \mathrm{d}E_p \, \Phi_p(E_p, \vec{r}) \times  \nonumber \\  
    \times \left( \frac{\mathrm{d} \sigma_{pp}}{\mathrm{d}E} + f_\mathrm{He} \, \frac{\mathrm{d}\sigma_{pHe}}{\mathrm{d}E} \right) \, .
\end{eqnarray}
Here, $\Phi_p$ denotes the CR proton flux calculated with CRPropa at a given position $\vec{r}$ in the Milky Way, $n_H$ is the target H-I gas density from the default ring model defined in HERMES and $f_{He} = 0.1$ is the fraction of target helium in the ISM. For the differential cross section $\mathrm{d}\sigma/\mathrm{d}E$, the AAfrag model \citep{AAfrag} is used.
To infer the impact of the anisotropy of the diffusion tensor, two different simulations are performed. In the first one, an isotropic diffusion is used. In the second simulation, we assume a ratio $\epsilon = 10^{-2}$ between the diffusion perpendicular and parallel to the field line. As the background magnetic field, the solenoidal improved version of the JF12 model is used \citep{JF12_sol}. To fill the central gap in the model, we superimpose the magnetic field of the CMZ defined in \cite{Guenduez2020ACenter}. Figure \ref{fig:field} shows a face-on and edge-on view of the field lines in this field geometry. 
The spatial source distribution follows the distribution inferred by \cite{BlasiAmato2012a}. 
From the individual $\gamma$-ray emission maps the relative difference is calculated and shown in Fig.\ \ref{fig:skymap}. The regions colored in red correspond to higher $\gamma$-ray fluxes for the isotropic diffusion, and the blue regions are increased for the anisotropic diffusion. In general, the all-sky difference is divided into four main regions by the inner and outer Galaxy and the Galactic plane and halo. 

In the outer Galactic plane, the magnetic field lines are mainly oriented in the plane. Therefore, the CRs are more confined in the anisotropic diffusion, leading to higher $\gamma$-ray emissions in this region. Going towards the Galactic Center, the impact of the x-field structure is more important, and the field lines become more perpendicular to the plane. Therefore, the anisotropic diffusion allows the particles to escape faster from this region. Subsequently, the $\gamma$-ray emission in the isotropic diffusion is higher. 
To discuss the $\gamma$-ray emission in the Galactic halo, the question about the transport from the sources, which are lying in the plane, towards the region of interest is more important than the local confinement of CRs. Therefore, the structure of the inner and outer parts of the Milky Way is reversed. In the outer part of the halo the $\gamma$-ray emission is higher for the isotropic diffusion. In this case, the CRs can escape from the Galactic plane via perpendicular diffusion. In the inner part, the parallel transport is more efficient, resulting in higher $\gamma$-ray fluxes for the anisotropic diffusion. Here, one can also see the differences between the northern and southern halo, which originates from the differences in the magnetic field structure (compare Fig.\ \ref{fig:field}). The magnetic field lines in the northern hemisphere show a clear connection towards the poles, and therefore, the anisotropic $\gamma$-ray emission is dominant at all latitudes. In the southern hemisphere, the field lines are more twisted, and only in the halo closer to the galactic plane does the anisotropic diffusion lead to the higher $\gamma$-ray fluxes. At the lowest latitudes, the isotropic diffusion leads to higher $\gamma$-ray production as the CRs can enter this region only by perpendicular diffusion.
In general, the all-sky gamma ray is sensitive to the anisotropy in the CR diffusion and the magnetic field structure. In future work, a direct comparison of the predictions from the anisotropic CR simulations with all-sky observations by Fermi-LAT or LHASSO might help to constrain the anisotropy of the diffusion tensor. Due to the high correlation with the magnetic field structure, it might even become possible to test Galactic Magnetic Field models with $\gamma$-ray observations. 

\vspace{-0.5cm}

\section{Gamma-rays from active Galaxies: ballistic VS diffusive propagation}

\begin{figure}[t]
    \centering
    \includegraphics[width=\linewidth]{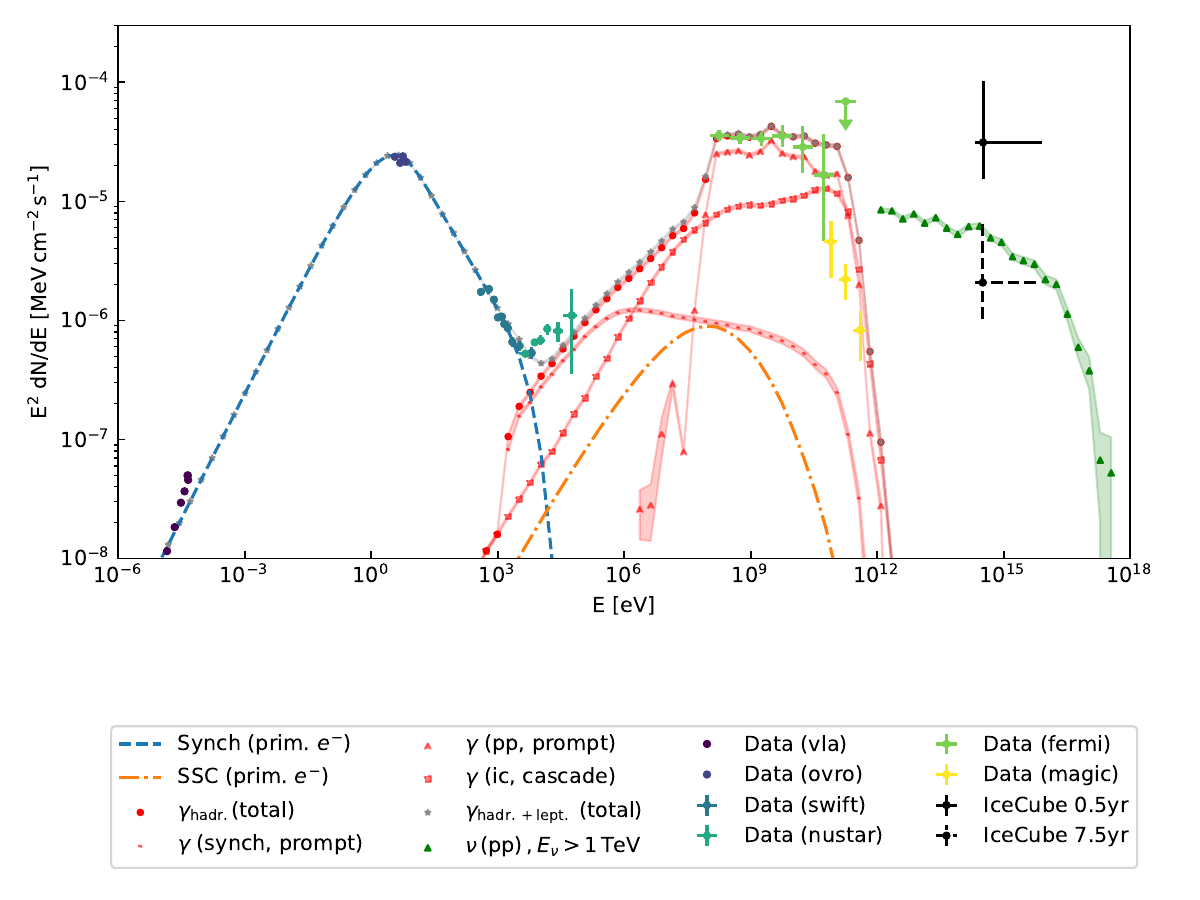}
    \caption{Multimessenger fit for TXS0506+056 for the 2017 multimessenger flare.}
    \label{fig:txs}
\end{figure}

The class of active galaxies has long been discussed to be among the few candidate classes for the acceleration of cosmic rays (CRs) up to the highest energies of $10^{20}$~eV. Yet, a consistent scheme for a mechanism to accelerate particles from thermal equilibrium to the highest energies is still to be found. Recent observations by the IceCube Neutrino Observatory indicate the emission of neutrinos from two types of active galaxies, Seyfert galaxies (with emission dominated by a bright core) and blazars (objects dominated by a jet that is pointing toward Earth).
The Seyfert galaxy NGC1068 shows evidence for neutrino emission \citep{ngc1068_icecube2022}. While in hadronic interactions, $\gamma$-rays and neutrinos are co-produced with about equal luminosities \citep{becker2008}, the detected $\gamma$-ray luminosity is orders of magnitude lower than the one of the neutrinos. The hadronic interactions must therefore happen in a $\gamma$-ray absorbing environment. The Corona close to the supermassive black hole
in the core of the active galaxy is well-suited as the origin, as the X-ray field of the Corona would highly absorb the $\gamma$-rays \citep{murase2020}, in fact so strong that the Fermi detection of NGC1068 in the GeV range must come from the surrounding starburst \citep{eichmann2016,eichmann2022}. The small jet of NGC1068 does not contribute significantly to the high-energy signatures \citep{salvatore2024}.
Further evidence comes from blazars, in particular, the source TXS0506+056. This source shows a $>3\,\sigma$ significance in the point source sample of IceCube \citep{ngc1068_icecube2022}, but there is also evidence for specific times of enhanced emission \citep{a7:txs_science2018}. The detection of the different emissions from TXS0506+056 are not easy to explain, as an early flare in 2014/2015, consisting of a larger number of neutrinos with around $\sim 10$~TeV energies are produced in a gamma-quiet state, while the high-energy neutrino (290~TeV) detected in 2017 comes at a time of high $\gamma$-ray activity. The 2017 neutrino flare is often explained by p$\gamma$ interactions in the jet \citep[e.g.][]{a7:xue2019,a7:wang2018,a7:reimer2019}, which produces a neutrino peak at hundreds of TeV energies that is not visible in $\gamma$-rays, as the $\gamma$-ray instruments do not cover the highest energies at the needed sensitivities, and additionally, considering the source redshifts, the photons typically undergo significant absorption in the extragalactic background light. The $\gamma$-rays at GeV energies as observed with Fermi are instead explained with Figure \ref{fig:txs} shows a fit to the data of the 2017 flare, considering two weeks of the $\gamma$-ray flare that is centered around the neutrino. The fit is done with a modified version of CRPropa as published in \citep{hoerbe2020}. Here, the time evolution and the SED of the flare are fit simultaneously. This procedure implies constraints in time \textit{and} in space for the fitting routine and is a path that will make it possible in the future to systematically distinguish between different emission scenarios. 

The CRPropa environment allows for 3D propagation, in particular with the option to propagate particles with the equation of motion approach and thus without explicitly assuming a diffusion tensor. This significantly reduces the uncertainty in the modeling. It also allows to naturally catch the transition from a diffusive to a ballistic transport behavior. Above a reduced rigidity of $\rho = r_g/l_c \gtrsim  5/(2\pi)$ \citep{10.1093/mnras/staa2533}, particles cannot scatter resonantly with all turbulent fluctuations. This changes the propagation effects in terms of the energy behavior of the diffusion tensor and the particles enter the so-called quasi-ballistic regime. Using the relativistic gyroradius $r_g=E/(q\cdot B \cdot c)$, the equation above provides a useful limit to the energy at which a particle should be treated ballistic, due to less and less resonant interactions \citep{10.1093/mnras/staa2533}, which is reached for energies \citep{2022Physi...4..473B}
\begin{equation}
    E\gtrsim Z \cdot \left( \frac{l_c}{10^{11}\,\mbox{m}}\right) \cdot \left(\frac{B}{0.42\,\mbox{G}}\right)\cdot 10^{15}\,\mbox{eV}\,.
    \label{eq:for_Hillas_plot}
\end{equation}
Here, the component of the diffusion tensor parallel to the ordered magnetic field changes its energy behavior from $D\propto E^{1/3}$ (Kolmogorov) to   $D\propto E^{2}$. At the highest energies, particles become ballistic \citep{10.1093/mnras/staa2533}, specifically at 
\begin{equation}
    \frac{E}{\mathrm{PeV}} \gtrsim Z\cdot \left(\frac{B}{3.3\,\mathrm{mG}}\right) \cdot \sqrt{\left(\frac{R}{10^{14}\,\mathrm{m}}\right)\cdot \left(\frac{l_c}{10^{12}\,\mathrm{m}}\right)}\,,
    \label{eq:for_Hillas_plot_2}
\end{equation}
with $R$ as the blob radius.
When knowing the  magnetic field strength $B$, magnetic coherence length $l_c$, and particle energy $E$, Eqns.\ (\ref{eq:for_Hillas_plot}) and (\ref{eq:for_Hillas_plot_2}) can be used to estimate in what regimes the particles are propagating in typical astrophysical sources of CRs
\citep{2022MNRAS.514.2658R}.

\begin{figure}[t]
    \centering
    \includegraphics[width=0.8\linewidth]{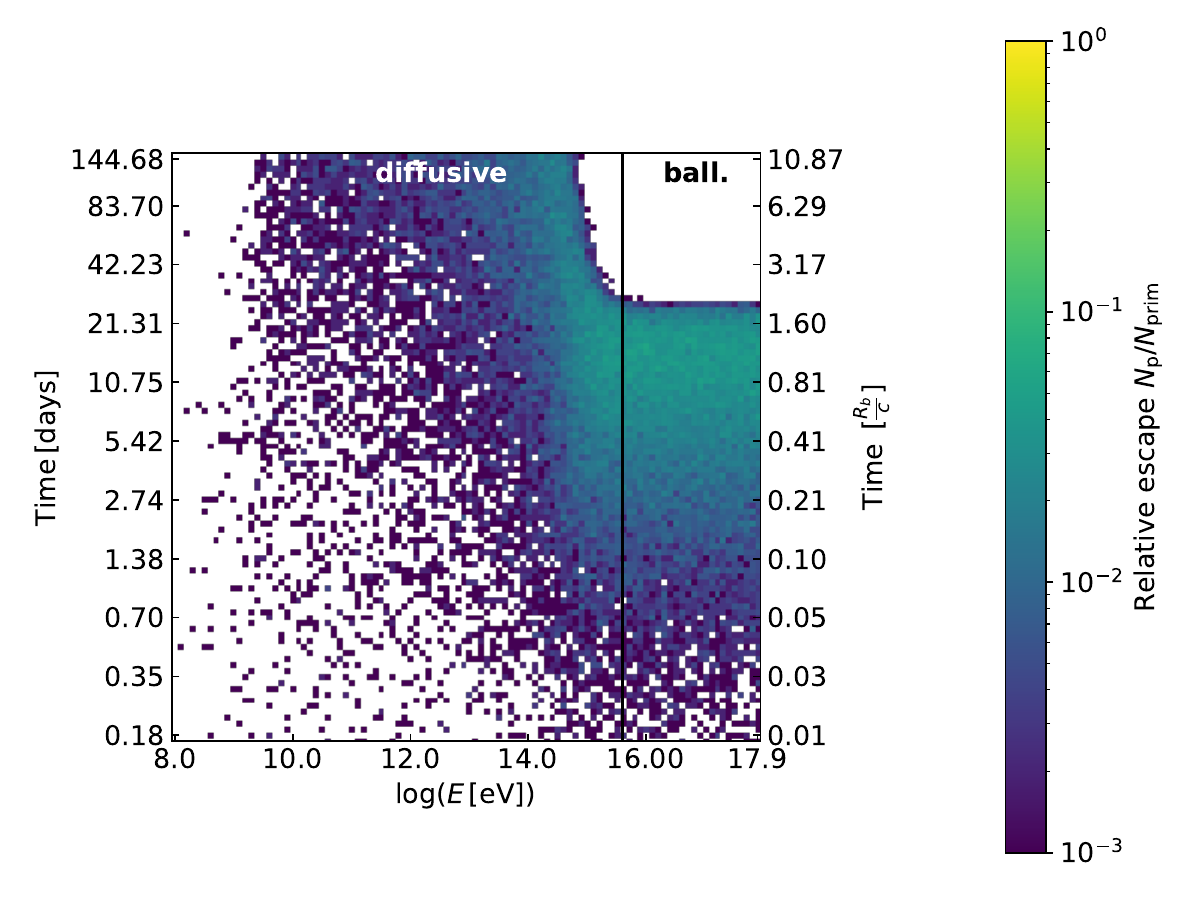}
    \caption{Number of particles arriving at the surface of the plasmoid with a radius of $R=10^{14.5}$~m for particles propagating in a turbulent magnetic field of strength $4\,\mathrm{mG}$. }
    \label{fig:diffusive_ballistic}
\end{figure}

Figure \ref{fig:diffusive_ballistic} shows the simulation result studying for the foregoing example of the TXS0506+056 fit only the transport behavior. Here, in the source frame of reference, particles are injected isotropically in a plasmoid of radius $R=10^{14.5}$~m 
in a turbulent magnetic field of the strength $4$~mG. The coherence length is assumed to be $l_c=0.01\cdot R= 10^{12.5}$~m. According to Eqns.\ (\ref{eq:for_Hillas_plot}) and (\ref{eq:for_Hillas_plot_2}), the transition from resonant scattering to the quasi-ballistic regime should happen at around $300\,\mathrm{TeV}$ primary energy (corresponding to $\sim 30$~TeV secondary photon and neutrino energy) 
and further to the purely ballistic propagation at around $4\,\mathrm{PeV}$ (see black vertical line in Fig.\,\ref{fig:diffusive_ballistic}). 
While the first transition is not visible due to a maximal trajectory length the particles were allowed to travel in the simulation (resulting in the cut of the figure at around 145 days), the second transition can be seen clearly in the plot. Here, the number of particles that reach the surface of the blob at a given time and with a given energy is shown. Before the transition, the distribution falls with a distinct slope, transitioning towards an energy-independent behavior once the particles propagate ballistically. This effect is expected to fold into the SED as well \citep{2022Physi...4..473B} and to change the observed spectrum at
\begin{equation}
    E_{\mathrm{transition}}^{\rm lab} = \left(\frac{\delta}{2}\right)\,\left(\frac{1+0.34}{1+z}\right)\,0.3 \rm{PeV} \, ,
\end{equation} where a Doppler factor of $\delta=2$ and a source redshift of $z=0.34$ \citep{Paiano_2018} is assumed. This energy translates into a $\gamma$-ray energy of $\sim 30$~TeV and thus above the cutoff of the spectrum due to EBL absorption. In the neutrino spectrum, the transition is expected at $\sim 20$~TeV, something that is not visible due to the cut of the simulation at early times. The second break is expected at $\sim 300$~TeV, where the change in the spectrum is observed (Fig.\ \ref{fig:txs}).

%
%




\vspace{-0.2cm}
 
\begin{acknowledgements}
We acknowledge support from the BMBF, grant no.\ 05A23PC3, and from the DFG grants via project 437789084 ("Modeling the Time-Dependent Cosmic-Ray Sun Shadow and its related Gamma-Ray and Neutrino Signatures") and via SFB1491 "Cosmic Interacting Matters - from Source to Signal" (project no.\ 445052434). 
\end{acknowledgements}

\bibliographystyle{aa}
\bibliography{bibliography}

\end{document}